\documentclass[manuscript]{aastex}
\begin{document}
\def\be{\begin{equation}}
\def\ee{\end{equation}}
\def\arccot{\rm arccot }
\def\bdm{\begin{displaymath}}\def\edm{\end{displaymath}}
\def\kr{cosmic ray }
\def\krs{cosmic rays }
\def\pr{\parallel }
\def\fp{Fokker-Planck }
\def\l{\left}
\def\r{\right}
\def\vpa{v_{\parallel }}
\def\vper{v_{\perp }}
\def\kpa{k_{\parallel }}
\def\kper{k_{\perp }}
\def\a{\alpha }
\def\ea{\rm{sgn} (q_a)}
\shorttitle{Cosmic ray transport}

\title{Cosmic ray transport theory in partially turbulent space plasmas with 
compressible magnetic turbulence}
\author{S. Casanova$^{1}$, R. Schlickeiser$^{1,2}$}
\affil{1 Institut f\"ur Theoretische Physik, Lehrstuhl IV:
Weltraum- und Astrophysik, Ruhr-Universit\"at Bochum,
D-44780 Bochum, Germany\\
2 Research Department Plasmas with Complex Interactions, Ruhr-Universit\"at Bochum, D-44780 Bochum, Germany}
\email{sabrina@tp4.rub.de}
\begin{abstract}

Recently a new transport theory of \krs in magnetized space plasmas extending 
the quasilinear approximation to the particle orbit has been developed 
for the case of an axisymmetric incompressible magnetic turbulence. Here we 
generalize the approach to the important physical case of a 
compressible plasma. As previously obtained in the case of an 
incompressible plasma we allow arbitrary gyrophase deviations 
from the unperturbed spiral orbits in the uniform magnetic field. 
For the case of quasi-stationary and spatially homogeneous magnetic turbulence 
we derive in the small Larmor radius approximation gyro-phase averaged 
\kr \fp coefficients. Upper limits for the perpendicular and pitch-angle 
\fp coefficients and for the perpendicular and parallel spatial diffusion 
coefficients are presented.
\end{abstract}


\section{ Introduction}

The study of the \kr transport in turbulent magnetic fields is crucial in many aspects of high energy astrophysics, 
such as the efficiency of \kr diffusive 
shock acceleration, the modulation and penetration of low energy \krs in the heliosphere and their 
confinement and escape from the Galaxy.

A new theory of cosmic ray transport in magnetized plasmas extending 
the quasilinear approximation to the particle orbit has been recently published by one of us (Schlickeiser, 2011) (~hereafter Paper 1~). 
In Paper 1 the transport parameters of energetic charged particles in turbulent magnetized cosmic 
plasmas were derived for the case of an incompressible plasma, i.e. 
plasmas 
for which the component of the magnetic turbulence, $\delta B_z=0$, parallel to the guide magnetic field, $\vec{B_0}=B_0\vec{e}_{z}$, is set to zero. Here we present 
the generalization of the theory to the case of compressible magnetic turbulence with $\delta B_z\ne 0$.

In Section 2 we briefly review the theory developed in Paper 1. In Section 3 
we obtain the gyro-phase averaged 
\kr \fp coefficients for a 
quasi-stationary, spatially homogenous turbulence under a Corrsin-type assumption 
on the nature of generalized orbits (Corrsin 1959, Salu and Montgomery, McComb 1990). 
Simplified formulas for the gyro-phase averaged \kr \fp coefficients are obtained in Section 4 assuming that the magnetic 
turbulence is asymmetric, while 
the quasilinear limit of the coefficients is shown in the Appendix. In Section 5 from the \fp coefficients we derive 
upper and lower limits for the perpendicular and parallel spatial diffusion coefficients. In Section 6 we compare the relative importance 
of mirror forces and turbulent scattering for the cosmic ray transport in interstellar plasmas.

\section{The gyro-averaged Fokker-Planck equations}
\subsection{Equations of motion of a particle in magnetic fields}
For the following treatment we remind shortly the equation of motion of charged particles of mass $m$, charge $q$, 
and Lorentz factor $\gamma =\sqrt{1+(p/mc)^2}$ in a uniform guide magnetic field $\vec{B}_0=B_0\vec{e}_z=(0,0,B_0)$. A random magnetic field, 
$\delta \vec{B}$, is 
superposed to the guide field. 

\be
\dot {\vec{p}}={q\over m\gamma c}\vec{p}\times \l[\vec{B_0}+\delta \vec{B}\r],\;\; \dot {\vec{x}}=\vec{v}={\vec{p}\over \gamma m}
\label{aa3}
\ee 
The scalar product of Eq. (\ref{aa3}) with $\vec{p}$ readily yields $p=|\vec{p}|=$const., $v=$const. 
and $\gamma =$const.. Introducing the constant relativistic gyrofrequency $\Omega ={qB_0\over \gamma mc}$ 
and scaling the turbulent fields in units of $B_0$, $\delta \vec{b}={\vec{B}\over B_0}$ we obtain 

\be
{d\over dt}\left( \matrix{v_x\cr v_y \cr v_z \cr}\right)=
\Omega \left( \matrix{ v_y(1+\delta b_z)-v_z\delta b_y \cr 
                       -v_x(1+\delta b_z)+v_z\delta b_x \cr 
		       v_x\delta b_y-v_y\delta b_x \cr }\right)
\label{aa6}
\ee

For the time evolution of the particle pitch-angle cosine $\mu =v_z/v$ and phase $\phi =\arctan (v_y/v_x)$ this implies 

\be
{d\mu \over dt}=h_{\mu }(t)={\Omega \over v}\l(v_x\delta b_y-v_y\delta b_x\r)=\Omega \sqrt{1-\mu ^2}\l(\cos \phi \; \delta b_y-\sin \phi \; \delta b_x\r),
\label{aa7}
\ee
and 

\be
{d\phi \over dt}=-\Omega +h_{\phi }(t),\;\; h_{\phi }(t)=-\Omega \delta b_z+{\Omega \mu \over \sqrt{1-\mu ^2}}\l(\cos \phi \; \delta b_x+\sin \phi \; \delta b_y\r).
\label{aa8}
\ee
with the two random forces $h_{\mu }(t)$ and $h_{\phi }(t)$.

In the coordinates of the guiding center
\be
\vec{X}=(X,Y,Z)=\vec{x}+{\vec {v}\times \vec{e}_z\over \Omega }=\vec{x}+{1\over \Omega }\left( \matrix{v_y\cr -v_x \cr 0 \cr}\right)
\label{a1}
\ee

 Eqs. (\ref{aa6}) become

\be
{d\over dt}\left( \matrix{X \cr Y \cr Z \cr}\right)=
\left( \matrix{ v_z\delta b_x-v_x\delta b_z \cr  v_z\delta b_y-v_y\delta b_z \cr v_z \cr }\right).
\label{aa9}
\ee

Indicating 
$X_i=[X,Y]$ with $i,j=1,2$,  Eq.(\ref{aa9}) provide the two additional random force terms $h_{i}(t)$, proportional to the turbulent magnetic field components

\be
{dX_i\over dt}=h_i(t)=v_z(t)\delta b_i(t)-v_i(t)\delta b_z(t),
\label{aa10}
\ee

\subsection{The ensemble-averaged particle distribution function}

The description of the \kr transport within a large-scale guide magnetic 
field, which is uniform on the scales of the \kr particles gyradii 
$R_L=v/|\Omega |$, 
is given by the solution of the Vlasov (collision-free Boltzmann) equation 
for the particle distribution function $F$ 
(Hall and Sturrock 1968, Schlickeiser 2002). In spherical momentum coordinates $(X,Y,z,p,\mu ,\phi )$ the Vlasov equation reads 
(Hall and Sturrock 1968, Achatz et al. 1991)

\be
{\partial F\over \partial t}+\; v\mu {\partial F\over \partial z}-\Omega {\partial F\over \partial \phi }+
p^{-2}{\partial\over \partial p}\l[p^2h_{p}(t)F\r]+{\partial\over \partial y_{\a}}\l[h_{\a }(t)F\r]-Q_0(z,X,Y,p,\mu , \phi ,t)=0,
\label{aa1}
\ee

where $y_{\a }\in [X,Y, \mu ,\phi ]$ and 

\be
Q_0(z,X,Y,p,\mu ,\phi ,t)=S_0(z,X,Y,p,\mu ,\phi ,t)-{\cal N}_0F-{\cal R}_0F
\label{aa2}
\ee
accounts for sources and sinks ($S_0$) and the effects of the mirror force (${\cal N}_0$) and momentum loss processes 
(${\cal R}_0$), 
where the latter two operate on much longer spatial and time scales than the particle interactions with the stochastic 
fields. In Eq. (\ref{aa1}) we use the Einstein sum convention for indices and the 
short notation $\partial _{\nu }=(\partial/\partial x_{\nu })$. $x_{\nu ,\sigma }\in [\mu ,p, X,y]$ 
represent the four phase space variables $\mu ,p,X,Y$ with non-vanishing stochastic fields $h_{\nu }(t)$. 

The particle distribution function, $F(X,Y,z,p,\mu ,\phi )$, varies in a irregular way under the influence of the stochastically 
fluctuating fields, $h_{\nu }(t)$. However, we do not look for the detailed function $F$, but rather we 
look for an ensemble-averaged solution, $<F>$, an expectation value of of Equ.~\ref{aa1}, which results from averaging over different 
realizations of the fields $h_{\nu }(t)$ with 
the same statistical averages. In the following treatment we will keep only first-order terms in the fluctuating quantities $\delta F$
\be
F = < F > + \delta F
\ee
and in the turbulent fields, $h_{\nu }(t)$. In other words we will consider the case of weak turbulence or 
quasilinear approximation. Moreover we will also neglect electric fields ($h_p =0$). As shown in details in Paper 1, under the assumption of weak turbulence 
the ensemble-averaged solution, $<F>$, can be obtained by solving the kinetic equation 

\bdm
\partial _t<F>\; +v\mu \partial _z<F>\; -\Omega \partial _{\phi }<F>-Q_0(z,X,Y,p,\mu , \phi ,t)=
-{\partial\over \partial x_{\a}}P_{\a \sigma }{\partial <F>\over \partial x_{\sigma }}
\edm
\be
-{\partial\over \partial \phi }P_{\phi \sigma }{\partial <F>\over \partial x_{\sigma }}\; -
{\partial\over \partial x_{\a}}P_{\a \sigma }{\partial <F>\over \partial \phi }
\label{ad13}
\ee

where $x_{\nu }\in [X,Y,\mu ]$ and the \fp coefficients are given as

\be
P_{\a \sigma }=<h_{\a }(t)\int_{t_0}^tds\, h_{\sigma }(s)>
\label{ad12}
\ee

The time-integration operator in Equ.~\ref{ad12} is performed over a generalization of the unperturbed gyrocenter orbit in the uniform magnetic field 
with deviations of the gyrophase given by 
\be
X_s=X,\; Y_s=Y,\; Z_s=Z+v\mu (s-t),\; p_s=p,\; \mu _s=\mu ,\; \phi _s=\phi -\Omega (s-t)+\delta \phi (t-s),
\label{ad8}
\ee
that contains the additional arbitrary gyrophase variation $\delta \phi (t-s)$, with $\delta \phi =0$ for $s=t$.

Fourier transforming the stochastic force in space, the time integral in Equ.~\ref{ad12} becomes

\bdm
\int_{t_0}^tds\, h_{\sigma }(s)=\int d^3k\, \int_{t_0}^tdu\, H_{\sigma }(\vec{k},s)
\edm
\be
\times
\exp \l[\imath \vec{k}\cdot \vec{X}
+\imath v\mu \kpa (s-t) +\imath \kper v\sqrt{1-\mu ^2}\int^sdw\, \cos \l(\psi -\phi +\Omega (w-t)-\delta \phi (t-w)\r)\r]
\label{ad11}
\ee

where the particle position is given as

\be
\vec{x}(s)=\left( \matrix{ X+v\sqrt{1-\mu ^2}\int^sdw\, \cos (\phi -\Omega (w-t)+\delta \phi (t-w)) \cr 
                           Y+v\sqrt{1-\mu ^2}\int^sdw\, \sin (\phi -\Omega (w-t)+\delta \phi (t-w)) \cr 
			   Z+v\mu (s-t)\cr }\right) 
\label{ad10}
\ee
and where we have introduced cylindrical coordinates for the wavenumber vector $\vec{k}=(k_\perp \cos \psi ,k_\perp \sin \psi, \kpa )$ and the 
particle velocity.

As explained in details in Paper 1, in the {\it small Larmor radius approximation} (Chew et al. 1956, Kennel and Engelmann 1962) 
the distribution functions are independent of $\phi $ to lowest order and can then be expanded as

\be
<F>=F_0+{F_1\over \Omega }
\label{ac1}
\ee

and the Larmor-phase-averaged equation becomes

\be
\partial _tF_0+v\mu \partial _zF_0-Q(z,X,Y,p,\mu ,t)=
-{\partial\over \partial x_{\a}}D_{\a \sigma }{\partial F_0\over \partial x_{\sigma }}-
\label{ac4}
\ee

with the gyro-averaged source term

\be
Q(z,X,Y,p,\mu ,t)={1\over 2\pi }\int_0^{2\pi }d\phi \, Q_0(z,X,Y,p,\mu , \phi ,t),
\label{ac5}
\ee

and the gyro-averaged \fp coefficients 

\be
D_{\a \sigma }=\Re {1\over 2\pi }\int_0^{2\pi }d\phi P_{\a \sigma }=
\Re {1\over 2\pi }\int_0^{2\pi }d\phi <h_{\a }(t)\int_{t_0}^tds\, h^*_{\sigma }(s)>,
\label{ac6}
\ee

where we replaced $h_{\sigma }(t)=h^*_{\sigma }(t)$ by its complex conjugate because the stochastic forces are real-valued quantities. 

The generalization of the time integral in Eq.~\ref{ad12} from the unperturbed motion of the gyrocenters in the guide magnetic 
field to arbitrary gyrophase motions of particles is possible essentially because of the 
gyrophase-averaging in Eq.~\ref{ac6}. As demonstrated in Paper 1 the considered general particle gyrophase motion then only 
modifies the arguments of trigonometric and 
Bessel functions as compared to the quasilinear approximation of 
particle orbits.

\section{Derivation of \fp coefficients for compressible magnetic turbulence}

Following the approach of Paper 1 we now derive the \fp coefficients for the case of 
compressible magnetic turbulence. We make the following assumptions on the nature of the turbulence: 
the turbulence is quasi-stationary, meaning that the correlation function 
$<h^*_{\nu }(t)h_{\sigma }(s)>$ depends only on the absolute value of the time difference 
$|t-s|=|\tau |$, so that with the substitution $s=t-\tau $ we find for Eq.(\ref{ac6}) 

\be
D_{\nu \sigma }=\Re {1\over 2\pi }\int_0^{2\pi }d\phi \int_{t_0}^{t}ds\, <h_{\nu }(t)h^*_{\sigma }(s)>
=\Re {1\over 2\pi }\int_0^{2\pi }d\phi \int_0^{t-t_0}ds\, <h_{\nu }(t)h^*_{\sigma }(t-\tau )>
\label{b3}
\ee

As second assumption we use that the turbulent magnetic fields are homogenously distributed, 
meaning that independent from the actual position of the gyrocenter at time $t$ the particles are subject to 
turbulence realizations with equal statistical properties. This allows us to average the \fp coefficients over the spatial 
position of the guiding center using
\be
\frac{1}{(2\pi)^{3}}\int_{-\infty}^{\infty}d^3X\; e^{i(\vec{k}^{'}-\vec{k})\cdot \vec{X}}=\delta(\vec{k}^{'}-\vec{k}),
\label{b8}
\ee
implying that turbulence fields at different wavevectors are uncorrelated. As explained in details in Paper 1, the \fp coefficients then become 

\bdm
D_{\nu \sigma }=\Re {1\over 2\pi }\int_0^{2\pi }d\phi \, \int d^3k\, 
\int_0^{t-t_0}d\tau \, <H_{\nu  }(\vec{k},t)H_{\sigma }^*(\vec{k},t-\tau )e^{\imath v\mu \kpa \tau }
\edm
\be
\times \exp \l[-\imath \kper v\sqrt{1-\mu ^2}\l(\int^{t-\tau }dw\, \cos (\psi -\phi +\Omega (w-t)-\delta \phi (t-w))+
{\sin (\phi -\psi )\over \Omega }\r)\r]>
\label{b9}
\ee

A third assumption concerns the nature of the particle orbits, i.e. we consider only orbits 
where $\delta \phi (w)$ does not depend upon the fluctuating fields, 
so that the ensemble averaging 
in Eq. (\ref{b9}) involves only the 2nd order correlation functions of the stochastic fields. 
This is generally called the Corrsin independence hypothesis (Corrsin 1959, Salu and Montgomery, McComb 1990). 

With $\xi =t-w$ and the abbreviation 
\be
G(\xi )=\Omega \xi +\delta \phi (\xi )
\label{b91}
\ee
the \fp coefficients (\ref{b9}) then are  

\bdm
D_{\nu \sigma }=\Re {1\over 2\pi }\int_0^{2\pi }d\phi \, \int d^3k\, 
\int_0^{t-t_0}d\tau \, <H_{\nu  }(\vec{k},t)H^*_{\sigma }(\vec{k},t-\tau )>
\edm
\be
\times \exp \l[\imath v\mu \kpa \tau +\imath \kper v\sqrt{1-\mu ^2}\l(\int^{\tau }d\xi \, \cos (\phi -\psi +G(\xi ))-
{\sin (\phi -\psi )\over \Omega }\r)\r],
\label{b10}
\ee

Later we will also assume that the turbulence has a finite decorrelation time $t_c$ such that the correlation functions  
$<h_{\nu }(t)h^*_{\sigma }(t-\tau )>\to 0$ fall to a negligible magnitude for $\tau \to \infty$, so that the upper 
integration boundary in the $\tau $-integral can be replaced by infinity

\be
D_{\nu \sigma }=\Re {1\over 2\pi }\int_0^{2\pi }d\phi \int_0^{\infty }d\tau \, <h_{\nu }(t)h^*_{\sigma }(t-\tau )>.
\label{b4}
\ee

We remark that diffusive transport of cosmic rays happens if the turbulence is quasi-stationary and has a finite decorrelation time $t_c$, 
because the resulting gyro-averaged \fp coefficients 
at large times $t-t_0\gg t_c$ no longer depend on the time-difference $t-t_0$.

The equations of motion of the guiding center (Eqs. \ref{aa10}) can be written as

\be
{dX_i\over dt}=h_i(t)=v\mu \delta b_i(t) -v_i(t) \delta b_z(t),
\label{b1}
\ee
if 

\be
v_i(t)= v \sqrt{1-\mu ^2} \cos((i-1)\frac{\pi}{2} - \phi )
\label{b1bis}
\ee

where $i=[1,2]$, whereas the pitch-angle random force (Eq. (\ref{aa7})) is

\be
{d\mu \over dt}=h_{\mu }(t)=\Omega \sqrt{1-\mu ^2}\l(\cos \phi \delta b_2(t)-\sin \phi \delta b_1(t)\r),
\label{b2}
\ee
\subsection{Individual gyro-averaged \fp coefficients}
The Fourier transforms of the stochastic fields in (\ref{b1}) and (\ref{b2}) are 

\bdm
H_i(\vec{k},t)=v\mu b_i(\vec{k},t) -  v_i(t) b_z(\vec{k},t),
\edm
\bdm
 H^*_i(\vec{k},t-\tau )=v\mu b_i^*(\vec{k},t-\tau )-  v_i(t-\tau) b_z^*(\vec{k},t-\tau),\; 
\edm
\bdm
H_{\mu }(\vec{k},t)=\Omega \sqrt{1-\mu ^2}\l(\cos (\phi )b_2(\vec{k},t)-\sin (\phi )b_1(\vec{k},t)\r)\; 
\edm
\be
H^*_{\mu }(\vec{k},t-\tau )=\Omega \sqrt{1-\mu ^2}\l(\cos (\phi +G(\tau ))b^*_2(\vec{k},t-\tau )-
\sin (\phi +G(\tau ))b^*_1(\vec{k},t-\tau )\r)
\label{b101}
\ee
where
\bdm
 v_i(t) = v \sqrt{1-\mu ^2} \cos \l( (i-1) \frac{\pi}{2} - \phi \r) ,\; v_i(t-\tau) =  v \sqrt{1-\mu ^2} \cos \l( (i-1) \frac{\pi}{2} - (\phi+G(\tau)) \r) \;
\edm

In terms of the magnetic field correlation tensor

\be
<b_i(\vec{k},t)b^*_j(\vec{k},t-\tau )>=P_{ij}(\vec{k}, \tau )
\label{b11}
\ee
we then obtain for the perpendicular \fp coefficients 

\bdm
D_{ij}=\Re {v^2 \over 2\pi }\int d^3k\, \int_{0}^{t-t_0}d\tau \, e^{\imath v\mu \kpa \tau }
\int_0^{2\pi }d\phi \,  \, H_{ij} (\vec{k}, \tau )
\edm \be \times
\exp \l[\imath \kper v\sqrt{1-\mu ^2}\l(\int^{\tau }d\xi \, \cos (\phi -\psi +G(\xi ))-
{\sin (\phi -\psi )\over \Omega }\r)\r],
\label{b12}
\ee
where
\bdm
 H_{ij}(\vec{k}, \tau ) =  {\mu}^2  P_{ij}(\vec{k}, \tau ) -  \mu \, \sqrt{1-\mu ^2}  \cos \l( (j-1) \frac{\pi}{2} - (\phi+G(\tau)) \r )  \, P_{iz}(\vec{k}, \tau )   \, + 
\edm
\bdm
- \,\mu  \, \sqrt{1-\mu ^2}  \cos \l( (i-1) \frac{\pi}{2} - \phi \r )  \, P_{zj}(\vec{k}, \tau )  \, + 
\edm
\be 
+ \, (1-\mu ^2)  \cos \l( (i-1) \frac{\pi}{2} - \phi \r) \cos \l( (j-1) \frac{\pi}{2} - (\phi+G(\tau)) \r) \, P_{zz}(\vec{k}, \tau )
\label{hij}
\ee
 
The mixed \fp coefficients are instead given as

\bdm
D_{i\mu }=\Re {v \Omega \sqrt{1-\mu ^2}\over 2\pi }\int d^3k\, \int_{0}^{t-t_0}d\tau \, e^{\imath v\mu \kpa \tau } \int_0^{2\pi }d\phi \,  \, H_{i\mu} (\vec{k}, \tau )
\edm
\be \times 
\exp \l[\imath \kper v\sqrt{1-\mu ^2}\l(\int^{\tau }d\xi \, \cos (\phi -\psi +G(\xi ))-
{\sin (\phi -\psi )\over \Omega }\r)\r],
\label{b13}
\ee
where
\bdm
 H_{i\mu} (\vec{k}, \tau ) =  \mu \cos \l(  \phi +G(\tau ) \r) \, P_{i2}(\vec{k}, \tau )-
\mu \sin \l( \phi +G(\tau ) \r) \, P_{i1}(\vec{k}, \tau ) \,+
\edm
\bdm
- \, \sqrt{1-\mu ^2} \cos \l((i-1) \frac{\pi}{2} -\phi \r) \cos(\phi+G(\tau)) P_{z2}(\vec{k}, \tau )  \,+
\edm
\bdm
+ \, \sqrt{1-\mu ^2} \cos \l((i-1) \frac{\pi}{2} -\phi \r) \sin(\phi+G(\tau)) P_{z1}(\vec{k}, \tau )
\edm

and

\bdm
D_{\mu i}=\Re {v \Omega \sqrt{1-\mu ^2}\over 2\pi }\int d^3k\, \int_{0}^{t-t_0}d\tau \, e^{\imath v\mu \kpa \tau }
\int_0^{2\pi }d\phi \, H_{\mu i} (\vec{k}, \tau )
\edm
\be
\times
\exp \l[\imath \kper v\sqrt{1-\mu ^2}\l(\int^{\tau }d\xi \, \cos (\phi -\psi +G(\xi ))-
{\sin (\phi -\psi )\over \Omega }\r)\r]  ,
\label{b14}
\ee
where
\bdm
H_{\mu i} (\vec{k}, \tau ) = \mu \cos \phi P_{2i}(\vec{k}, \tau )- \mu \sin \phi P_{1i}(\vec{k}, \tau )\,+
\edm
\bdm
- \, \sqrt{1-\mu ^2}  \cos \phi  \cos \l( (i-1) \frac{\pi}{2} -(\phi + G(\tau))  \r) P_{2z}(\vec{k}, \tau )\,+ 
\edm
\be
+ \, \sqrt{1-\mu ^2}  \sin \phi  \cos  \l( (i-1) \frac{\pi}{2} -(\phi + G(\tau)) \r) P_{1z}(\vec{k}, \tau ) 
\label{himu}
\ee

The \fp coefficients parallel to the direction of the guide magnetic field are 

\bdm
D_{\mu \mu }=\Re {\Omega ^2(1-\mu ^2)\over 2\pi }\int d^3k\, 
\int_{0}^{t-t_0}d\tau \, e^{\imath v\mu \kpa \tau }\int_0^{2\pi }d\phi \,H_{\mu \mu} (\vec{k}, \tau )
\edm
\be 
\times
\exp \l[\imath \kper v\sqrt{1-\mu ^2}\l(\int^{\tau }d\xi \, \cos (\phi -\psi +G(\xi ))-
{\sin (\phi -\psi )\over \Omega }\r)\r],
\label{b15}
\ee
where
\bdm 
H_{\mu \mu} (\vec{k}, \tau ) =\cos \phi \cos (\phi +G(\tau ))P_{22}(\vec{k}, \tau )+\sin \phi \sin (\phi +G(\tau ))P_{11}(\vec{k}, \tau )
\edm
\bdm
-\sin \phi \cos (\phi +G(\tau ))P_{12}(\vec{k}, \tau )-\cos \phi \sin (\phi +G(\tau ))P_{21}(\vec{k}, \tau ))
\edm

The $\phi $-integrals are calculated in Appendix A of Paper 1 and yield 
\be
D_{ij}=\Re  { v^2 \over 2  } \int d^3k\, \int_{0}^{t-t_0} d\tau \, e^{\imath v\mu \kpa \tau } I_{ij}(\vec{k}, \tau )
\ee
where 
\bdm
I_{ij}(\vec{k}, \tau ) = \l [ 2 {\mu}^2 P_{ij}(\vec{k}, \tau )  + (-1)^{j \mid i-j \mid }(1-\mu ^2) \cos \l( \mid i-j \mid \frac{\pi}{2} - G(\tau )\r)  P_{zz} (\vec{k}, 
\tau )  \r] J_0(Z) \, + 
\edm
\bdm
- \, 2 \imath \sqrt{1-\mu ^2} \mu   [ {(-1)}^{i-1} \sin \l( (i-1) \frac{\pi}{2} - 
( \psi  +\arcsin ({Z_1\over Z}) ) \r)  P_{zj}(\vec{k}, \tau )  + 
\edm
\bdm
{(-1)}^{j-1} \sin \l( (j-1) \frac{\pi}{2} - ( \psi  + G(\tau) +\arcsin ({Z_1\over Z})) \r)  
P_{iz}(\vec{k}, \tau ) ] J_1(Z)  \, + 
\edm
\bdm
 (-1)^{(i-1)(j-1)} (1-\mu ^2)  P_{zz} (\vec{k}, \tau )  
\cos \l( \mid i-j \mid \frac{\pi}{2} - ( 2 \psi  + G(\tau) +\arcsin ({Z_1\over Z}) ) \r) J_2(Z)
\edm
\be
\label{b17}
\ee
\be
D_{i\mu }=\Re { v \Omega \sqrt{1-\mu ^2}\over 2 }\int d^3k\, \int_{0}^{t-t_0}d\tau \, e^{\imath v\mu \kpa \tau } I_{i\mu}(\vec{k}, \tau ) 
\ee
where
\bdm
I_{i\mu}(\vec{k}, \tau ) =  \sqrt{1-\mu ^2} \l[ \sin \l( (i-1) \frac{\pi}{2} - G(\tau) \r) P_{z1}(\vec{k}, \tau ) 
+ {(-1)}^{i} \cos \l( (i-1) \frac{\pi}{2} - G(\tau) \r) 
P_{z2}(\vec{k}, \tau ) \r]  J_0(Z) 
\edm
\bdm
 + 2 \imath \mu \l[ \sin \l(\psi + G(\tau) + \arcsin ({Z_1\over Z})\r)P_{i2}(\vec{k}, \tau ) +
\cos \l(\psi + G(\tau) +\arcsin ({Z_1\over Z}) \r) P_{i1}(\vec{k}, \tau )\r] J_1(Z) \, + 
\edm
\bdm
 \, \sqrt{1-\mu ^2} \Big{[} {(-1)}^{i-1} \sin \l( (i-1) \frac{\pi}{2} - ( 2 \psi  + G(\tau) +
\arcsin ({Z_1\over Z})) \r)  P_{z1}(\vec{k}, \tau )  \, + 
\edm
\bdm
-  \cos \l( (i-1) \frac{\pi}{2} -( 2 \psi  + G(\tau) +\arcsin ({Z_1\over Z})) \r) 
P_{z2}(\vec{k}, \tau ) \Big{]}  J_2(Z)
\edm
\be
\label{b18}
\ee

\be
D_{\mu i}=\Re { v\mu \Omega \sqrt{1-\mu ^2}\over 2 }\int d^3k\, \int_{0}^{t-t_0}d\tau \, e^{\imath v\mu \kpa \tau }I_{\mu i}(\vec{k}, \tau ) 
\ee
where
\bdm
I_{\mu i}(\vec{k}, \tau )  = 
\sqrt{1-\mu ^2} \l[ {(-1)}^{i} \sin \l( (i-1) \frac{\pi}{2} - G (\tau) \r) P_{1z}(\vec{k}, \tau ) 
- \cos \l( (i-1) \frac{\pi}{2} - G(\tau) \r) P_{2z}(\vec{k}, \tau ) \r]  J_0(Z) 
\edm
\bdm
+  2 \imath \mu \l[\sin \l(\psi +\arcsin ({Z_1\over Z})\r)P_{2i}(\vec{k}, \tau )+
\cos \l(\psi +\arcsin ({Z_1\over Z})\r)P_{1i}(\vec{k}, \tau )\r]]J_1(Z) \, + 
\edm
\bdm
\sqrt{1-\mu ^2} \Bigl{[} {(-1)}^{i-1} \sin \l( (i-1) \frac{\pi}{2} 
-( 2 \psi  + G(\tau) +\arcsin ({Z_1\over Z})) \r) P_{1z}(\vec{k}, \tau ) \, + 
\edm
\be
- \, \cos \l( (i-1) \frac{\pi}{2} -( 2 \psi  + G(\tau) +\arcsin ({Z_1\over Z})) \r) P_{2z}(\vec{k}, 
\tau )  \Bigr{]}  J_2(Z) 
\label{b19}
\ee
and 
\bdm
D_{\mu \mu }=\Re {\Omega ^2(1-\mu ^2)\over 2}\int d^3k\, 
\int_{0}^{t-t_0}d\tau \, e^{\imath v\mu \kpa \tau }
\edm
\bdm 
\times
\Bigl[\l(\cos (G(\tau ))J_0(Z)-\cos (2\psi +G(\tau )+2\arcsin ({Z_1\over Z}))J_2(Z)\r)P_{11}(\vec{k},\tau )
\edm
\bdm
+\l(\cos (G(\tau ))J_0(Z)+\cos (2\psi +G(\tau )+2\arcsin ({Z_1\over Z}))J_2(Z)\r)P_{22}(\vec{k},\tau )
\edm
\bdm
-\l(\sin (G(\tau ))J_0(Z)+\sin (2\psi +G(\tau )+2\arcsin ({Z_1\over Z}))J_2(Z)\r)P_{21}(\vec{k},\tau )
\edm
\be
+\l(\sin (G(\tau ))J_0(Z)-\sin (2\psi +G(\tau )+2\arcsin ({Z_1\over Z}))J_2(Z)\r)P_{12}(\vec{k},\tau )\Bigr],
\label{b20}
\ee
respectively, where $J_n(Z)$ denotes the Bessel function of the first kind and order $n$,  

\be
Z_1=\kper v\sqrt{1-\mu ^2}\int^{\tau }d\xi \; \cos \l(G(\xi )\r)
\label{b21}
\ee
and 

\be
Z=\kper v\sqrt{1-\mu ^2}
\l[\l(\int^{\tau }d\xi \; \cos (G(\xi ))\r)^2+\l({1\over \Omega }+\int^{\tau }d\xi \; \sin (G(\xi ))\r)^2\r]^{1/2}
\label{b22}
\ee
\section{Axisymmetric turbulence}

Useful formulas can be obtained by assuming that the turbulence is asymmetric, meaning $P_{\alpha \beta}(\vec{k},\tau )$ are independent of the wave phase $\psi $

\be
P_{\alpha \beta}(\vec{k},\tau )=P_{\alpha \beta}(\kpa , \kper , \tau ),
\label{c1}
\ee

The integration over $\psi $ of the general formulas (\ref{b17}) - (\ref{b20}) then provides 
\bdm
D_{ij}=\Re \pi \, v^2 \int_{-\infty }^\infty d\kpa \, \int_0^\infty d\kper \, \kper \int_{0}^{t-t_0}d\tau \, e^{\imath v\mu \kpa \tau } \, J_0(Z)
\edm
\be
\times
\l [ 2 {\mu}^2 P_{ij}(\kpa , \kper , \tau ) + (-1)^{j \mid i-j \mid}(1-\mu ^2) \cos \l( \mid i-j \mid \frac{\pi}{2} - G(\tau )\r)  P_{zz} (\kpa , \kper , \tau )  \r ],
\label{c2}
\ee

\bdm
D_{i \mu }=\Re \, \pi v \Omega (1-\mu ^2) \int_{-\infty }^\infty d\kpa \, \int_0^\infty d\kper \, \kper 
\int_{0}^{t-t_0}d\tau \, e^{\imath v\mu \kpa \tau } \, J_0(Z)
\edm
\be
\times
\l[ \sin \l( (i-1) \frac{\pi}{2} - G(\tau) \r) P_{z1}( \kpa , \kper, \tau ) + {(-1)^i} \cos \l( (i-1) \frac{\pi}{2} - G(\tau) \r) P_{z2}(\kpa , \kper , \tau ) \r] 
\label{c22}
\ee

\bdm
D_{\mu i }=\Re \, \pi v \Omega (1-\mu ^2) \int_{-\infty }^\infty d\kpa \, \int_0^\infty d\kper \, \kper 
\int_{0}^{t-t_0}d\tau \, e^{\imath v\mu \kpa \tau } J_0(Z)
\edm
\be
\times
\l[ {(-1)}^{i} \sin \l( (i-1) \frac{\pi}{2} - G(\tau) \r) P_{1z}( \kpa , \kper , \tau ) -  \cos \l( (i-1) \frac{\pi}{2} - G(\tau) \r) P_{2z}( \kpa , \kper , \tau ) \r]
\label{c33}
\ee

and 

\bdm
D_{\mu \mu }=\Re \, \pi v^2\Omega ^2(1-\mu ^2) \int_{-\infty }^\infty d\kpa \, \int_0^\infty d\kper \, \kper 
\int_{0}^{t-t_0}d\tau \, e^{\imath v\mu \kpa \tau } \,J_0(Z)
\edm
\be
\times
\Bigl[\cos (G(\tau ))\l(P_{11}(\kpa , \kper ,\tau )+P_{22}(\kpa , \kper ,\tau )\r)
+\sin (G(\tau ))\l(P_{12}(\kpa , \kper ,\tau )-P_{21}(\kpa , \kper ,\tau )\r)\Bigr]
\label{c3}
\ee

Introducing the left-handed and right-handed polarized stochastic magnetic field components

\be
\delta b_{L,R}={1\over \sqrt{2}}\l[\delta b_1\pm \imath \delta b_2\r],
\label{c4}
\ee

so that 

\be
2P_{LL}=P_{11}+P_{22}+\imath P_{21}-\imath P_{12},\;\; 2P_{RR}=P_{11}+P_{22}+\imath P_{12}-\imath P_{21},
\label{c5}
\ee

we obtain for the pitch-angle \fp coefficient (\ref{c3}) 

\bdm
D_{\mu \mu }=\Re \, \pi \Omega ^2(1-\mu ^2) \int_{-\infty }^\infty d\kpa \, \int_0^\infty d\kper \, \kper 
\int_{0}^{t-t_0}d\tau \, J_0(Z)
\edm
\be
\times
\Bigl[e^{\imath (v\mu \kpa \tau +G(\tau )}P_{LL}(\kpa , \kper ,\tau )+e^{\imath (v\mu \kpa \tau -G(\tau )}P_{RR}(\kpa , \kper ,\tau )\Bigr]
\label{c6}
\ee

\section{Upper and lower limits of the general \fp coefficients in the diffusion limit}

If we now consider a magnetic field fluctuation decorrelation time $t_c=\gamma ^{-1}$ (Schlickeiser and Achatz 1993, Bieber et al. 1994)

\be
P_{ij}(\vec{k},\tau )=P^0_{ij}(\vec{k})e^{-\gamma \tau },
\label{n3}
\ee
then in the diffusion limit $t-t_0\gg t_c$ the general \fp coefficients (\ref{c2}) - (\ref{c6}) in asymmetric turbulence become

\bdm
D_{\mu \mu }=\pi \Omega ^2(1-\mu ^2) \int_{\infty }^\infty d\kpa \, \int_0^\infty d\kper \, \kper 
 \int_{0}^\infty d\tau \, J_0(Z)e^{-\gamma \tau }
 \edm
 \be
 \times
 \Bigl[\cos (v\mu \kpa \tau +G(\tau ))P^0_{LL}(\kpa , \kper )+\cos (v\mu \kpa \tau -G(\tau ))P^0_{RR}(\kpa , \kper )\Bigr]
 \label{e1}
 \ee
 and

 \bdm
 D_{ij}=  \pi v^2\int_{-\infty}^{\infty}d\kpa \int _0^\infty d\kper \, \kper \,  \int_{0}^\infty d\tau \, J_0(Z)  
\, e^{\imath v\mu \kpa \tau -\gamma \tau }
 \edm
 \be
\times
\l[  2  \, {\mu}^2 \, P^0_{ij}(\kpa , \kper ) + (1-\mu ^2) {1 \over 2} \l( e^{\imath G (\tau)} + e^ {- \imath G (\tau)} \r)  P^0_{zz} (\kpa , \kper ) \r]  
 \label{e21}
 \ee

\bdm
D_{i \mu }=  \pi v \Omega (1-\mu ^2) \int_{-\infty }^\infty d\kpa \, \int_0^\infty d\kper \, \kper \,  \int_{0}^\infty d\tau \, J_0(Z)  
\, e^{\imath v\mu \kpa \tau -\gamma \tau }
\edm
\be
\times
\l[ {1 \over 2} \l( e^{\imath G (\tau)} + 
e^ {-\imath G (\tau)} \r) P^0_{z1}( \kpa , \kper) {\delta}_{i2} - {1 \over 2} \l( e^{ \imath G (\tau)} + 
e^ {- \imath G (\tau)} \r) P^0_{z2}(\kpa , \kper) {\delta}_{i1}\r] 
\label{c22i}
\ee

\bdm
D_{\mu i }= \pi v \Omega \mu(1-\mu ^2) \int_{-\infty }^\infty d\kpa \, \int_0^\infty d\kper \, \kper \,  \int_{0}^\infty d\tau \, J_0(Z)  
\, e^{\imath v\mu \kpa \tau -\gamma \tau }
\edm
\be
\times
\l[ {1 \over 2} \l( e^{ \imath G (\tau)} + 
e^ {- \imath G (\tau)} \r) P^0_{1z}( \kpa , \kper ) {\delta}_{i2} - {1 \over 2} \l( e^{ \imath G (\tau)} + 
e^ {- \imath G (\tau)} \r) P^0_{2z}( \kpa , \kper ) {\delta}_{i1} \r]
\label{c33i}
\ee

Note that in Eqs (\ref{e1}-\ref{c33i}) we consider only the real part of the integral.

 Because of the existence of the finite turbulence decorrelation time $\gamma ^{-1}$, the correlation functions 
 $P_{ij}(\kpa , \kper ,\tau )$ fall to a negligible magnitude for $\tau \to \infty$, allowing us 
 to replace the upper integration boundary in the $\tau $-integrals in Eqs. (\ref{e1}), (\ref{e21}), (\ref{c22i}) and 
(\ref{c33i}) by infinity. 
 We recover the diffusion limit which is valid for times $t-t_0\gg \gamma ^{-1}$. 

If $J_0(A)\le 1$ and $\cos (x)\le 1$ Eqs. (\ref{e1} - \ref{c33i}) become

 \bdm
 D_{ij}<D^{\rm max}_{ij}={v^2\mu ^2 \over {\gamma} }\, 2\pi \int_{-\infty}^{\infty}d\kpa \int _0^\infty d\kper \, \kper P^0_{ij}(\kpa , \kper )  
\int_{0}^\infty d\tau \, J_0(Z) \cos(v\mu \kpa \tau)  
\, e^{-\gamma \tau }
\edm
\bdm
+ {v^2  {(1 - \mu^2 )} \over {2} }\, 2\pi  \int_{-\infty}^{\infty}d\kpa \int _0^\infty d\kper \, \kper P^0_{zz}(\kpa , \kper )   
\edm
\bdm
\int_{0}^\infty d\tau \, J_0(Z)  
\, \frac{e^{ -\gamma \tau }} {2}  \Bigl[ \cos \l( v\mu \kpa \tau  +G (\tau)  \r)  +   \cos \l( v\mu \kpa \tau -G (\tau) \r) \Bigr]  \delta_{ij} 
\edm
\be
 ={v^2\mu ^2\delta b^2_{ij}\over {\gamma} } + {v^2 {(1 - \mu^2 )}  \delta b^2_{zz} \delta_{ij} \over { 2 \gamma} }  
\label{e31}
 \ee
 and 

\bdm
D_{i \mu }<D^{\rm max}_{i\mu} = \frac{v \Omega (1-\mu ^2)}{2}\, 2\pi \int_{-\infty }
^\infty d\kpa \, \int_0^\infty d\kper \, \kper \int_{0}^\infty d\tau \, J_0(Z)  
\, \frac{e^{ -\gamma \tau }}{2} 
\edm
\bdm
\Bigl[ \l( cos \l( v\mu \kpa \tau  +G (\tau)  \r)  +   \cos \l( v\mu \kpa \tau -G (\tau) \r) \r)  \delta_{i2} P^0_{z1}( \kpa , \kper ) 
\edm
\bdm
-   \l[ cos \l( v\mu \kpa \tau  +G (\tau)  \r)  +   \cos \l( v\mu \kpa \tau -G (\tau) \r) \r]  
\delta_{i1} P^0_{z2}( \kpa , \kper ) \Bigr] 
\edm
\be
=  \frac{v \Omega (1-\mu ^2)}{2 \gamma }  
\l[\delta_{i2} \delta b^2_{z1} -   \delta_{i1} \delta b^2_{z2} \r]
\label{e32}
\ee

\bdm
D_{\mu i}<D^{\rm max}_{i\mu} = \frac{v \Omega (1-\mu ^2)}{2}\, 2\pi \int_{-\infty }^\infty d\kpa \, \int_0^\infty d\kper \, \kper \int_{0}^\infty d\tau \, J_0(Z)  
\, \frac{e^{ -\gamma \tau }}{2} 
\edm
\bdm
\Bigl[ \l[ \cos \l( v\mu \kpa \tau  +G (\tau)  \r)  +   \cos \l( v\mu \kpa \tau -G (\tau) \r) \r]  \delta_{i2} P^0_{1z}( \kpa , \kper ) 
\edm
\bdm
-   \l[ \cos \l( v\mu \kpa \tau  +G (\tau)  \r)  +   \cos \l( v\mu \kpa \tau -G (\tau) \r) \r]  \delta_{i1} P^0_{2z}( \kpa , \kper ) \Bigr]
\edm
\be
=  \frac{v \Omega (1-\mu ^2)}{2 \gamma }  \l[\delta_{i2} \delta b^2_{1z} -   \delta_{i1} \delta b^2_{2z} \r]
\label{e33}
\ee

 \bdm
 D_{\mu \mu }<D^{\rm max}_{\mu \mu }={\Omega ^2(1-\mu ^2)\over 2\gamma }2\pi \int_{-\infty}^{\infty}d\kpa \int _0^\infty d\kper \, \kper 
 \l[P^0_{LL}(\kpa , \kper )+P^0_{RR}(\kpa , \kper )\r]
 \edm
 \be
 ={\Omega ^2(1-\mu ^2)\over 2\gamma }\l[\delta b^2_{LL}+\delta b^2_{RR}\r]={\Omega ^2(1-\mu ^2)\over 2\gamma }\l[\delta b^2_{xx}+ \delta b^2_{yy}\r],
 \label{e4}
 \ee
 
where

 \be
 \delta b^2_{\mu\nu}=\int d^3kP^0_{\mu\nu}(\kpa , \kper )=2\pi \int_{-\infty}^{\infty}d\kpa \int _0^\infty d\kper \, \kper P^0_{\mu\nu}(\kpa , \kper )
 \label{e5}
 \ee

According to the diffusion approximation (Schlickeiser 2002), neglecting the influence of the mirror force contribution ${\cal N}_0$ in 
Eq. (\ref{aa1}), the perpendicular spatial diffusion coefficients for the isotropic part of the \kr phase space density are given by the pitch-angle average 

 \be
 \kappa _{ij}={1\over 2}\int_{-1}^1d\mu D_{ij}(\mu )
 \label{e6}
 \ee
 From Eq.~\ref{e31} we find the upper limits 

 \be
 \kappa _{ij}<\kappa ^{\rm max}_{ij}={v^2\over 3 \gamma } \l( {\delta b^2_{ij}} + \delta b^2_{zz} \delta_{ij} \r) 
 \label{e7}
 \ee
 
The parallel spatial diffusion coefficient for the isotropic part of the \kr phase space density is given by the pitch-angle average

 \be
 \kappa _{\parallel }={v^2\over 8}\int_{-1}^1d\mu {(1-\mu ^2)^2\over D_{\mu \mu }(\mu )}, 
 \label{e8}
 \ee

and its lower limit does not change with respect to the incompressible case examined in Paper 1.

 \be
 \kappa _{\parallel }>\kappa ^{\rm min} _{\parallel }={\gamma v^2\over 3\Omega ^2 \l[\delta b^2_{xx}+\delta b^2_{yy}\r]}
 \label{e9}
 \ee

\section{Mirror forces and turbulent scattering in the solar wind plasma}
 
Mirror forces are produced by large-scale spatial variations of the guide magnetic field. The perpendicular component of the mirror force generates 
gradients and curvature drifts of the cosmic ray guiding center ( Boyd and Sanderson 1969 ). In the presence of mirror forces the diffusion coefficients 
are given by the sum of the turbulent contribution, $k^{(T)}_{\mu\nu}$, and of the contribution due to the mirror forces, $k^{(M)}_{\mu\nu}$,

\be
\kappa_{\mu\nu} = \kappa^{(T)}_{\mu\nu} + \kappa^{(M)}_{\mu\nu}
\ee

We will now compare the effect of mirror forces and of the turbulent contribution, $k_{(\mu \nu )}^T$, calculated in Sect. 5, on the properties of cosmic ray transport in the solar wind plasma.

For mirror forces Schlickeiser and Jenko (2010) showed 
that in the case of a symmetric choice of the pitch-angle Fokker-Planck coefficients the ratio of the perpendicular
mirror spatial diffusion coefficient to the parallel turbulent spatial 
diffusion coefficient is given by the derivatives of the cosmic-ray Larmor radius. In particular, 
considering the case of a magnetic power spectrum of Alfvenic slab turbulence $P(k) \propto k^{-s}$ with $s<2$, 
Schlickeiser and Jenko 2010 obtained that the ratios of the non-zero perpendicular to parallel spatial diffusion coefficients are
 
\bdm
\frac{\kappa^{(M)}_{XX}}{\kappa^{(T)}_{ZZ}}  = \frac{2-s}{6-s} {(\frac{R_L}{3 L_2})}^2
\edm
\bdm
\frac{\kappa^{(M)}_{YY}}{\kappa^{(T)}_{ZZ}}  = \frac{2-s}{6-s} {(\frac{R_L}{3 L_1})}^2
\edm
\be
\label{71}
\ee

where we introduce the perpendicular magnetic field scale lengths (Schlickeiser and Jenko 2010)
\bdm 
{L_1}^{-1} = - B^{-1} \frac{\partial B}{\partial x} 
\edm
\be
{L_2}^{-1} = - B^{-1} \frac{\partial B}{\partial y} 
\label{7111}
\ee
 
Summing the diagonal terms of the diffusion matrix we obtain

\be
\frac{\kappa^{(M)}_{XX}+\kappa^{(M)}_{YY}}{\kappa^{(T)}_{ZZ}}  = \frac{2-s}{6-s} {(\frac{R_L}{3})}^2 \l( {1 
\over L^2_1} +{1 \over L^2_2} \r)
\label{7112}
\ee

where we remind that the \kr gyroradius $r_L$ is defined as 
\be
R_L = \frac{v}{\Omega} = \frac{p c}{Z e B_0} 
\label{72}
\ee

From Eq.~\ref{7112} we have 
\be
{\kappa}^{\rm min}_{\parallel } \, 
\l({\kappa^{(M)}_{XX}+\kappa^{(M)}_{YY}} \r) > v^2  {(\frac{R_L}{3})}^4 \, \frac{2-s}{6-s}  \l( {1 \over L^2_1} 
+{1 \over 
L^2_2} \r)
\label{7113}
\ee
since
\be
{\kappa}^{\rm min}_{\parallel } = {  v R_L \over 3} \,.
\ee
 
The relevant magnetic field random irregularities for the cosmic ray transport properties are the fast magnetosonic waves ( Lee and V\"olk 1975, Cho and Lazarian, 2003 ).
If we consider isotropic magnetosonic waves ( Schlickeiser 2002 )

\bdm
P_{xx} = { g(k) \over 8 \pi k^2} { \cos \Theta}^2
\edm
\bdm
P_{yy} = { g(k) \over 8 \pi k^2} 
\edm
\be
P_{zz} = { g(k) \over 8 \pi k^2} { \sin \Theta}^2
\label{e100}
\ee

then the random contributions to the field irregularities are  
\bdm
\delta b^2_{xx} = {1 \over 4} \int_{-1}^{1} d \mu \, {\mu}^2  \, \int dk g(k) = {1 \over 6}  \int dk g(k)
\edm
\bdm
\delta b^2_{yy} = {1 \over 4} \int_{-1}^{1} d \mu  \, \int dk g(k) = {1 \over 2} \int dk g(k)
\edm
\be 
\delta b^2_{zz} = {1 \over 4} \int_{-1}^{1} d \mu \, ( 1 - {\mu}^2) \, \int dk g(k) = {1 \over 3} \int dk g(k)
\ee

Using the upper and lower limits for the diffusion coefficients from random turbulent forces in Eqs. (\ref{e7}) and (\ref{e9}) and applying it for the case of fast magnetosonic waves we have  

\be
{\kappa^{(T)}}^{\rm min}_{\parallel } \l( {\kappa^{(T)}}^{\rm max}_{xx}+ {\kappa^{(T)}}^{\rm max}_{xx} \r ) \geq  {({v R_L\over 3})}^2  
\l[ 1  + { 2 \delta b^2_{zz}\over {\delta b^2_{xx}+\delta b^2_{yy}}} \r]  =  2 \,  {({v R_L\over 3})}^2 \,.
\label{e10}
\ee

Taking the ratio of Eq.~\ref{7113} with Eq.~\ref{e10} we obtain a relation for
the product of perpendicular diffusion coefficients independent of $\kappa _{ZZ}$
\begin{eqnarray*} 
\frac{ {\kappa^{(M)}_{XX}+\kappa^{(M)}_{YY}} } {{\kappa^{(T)}}^{\rm max}_{xx}+ {\kappa^{(T)}}^{\rm max}_{xx} } && >  
{(\frac{R_L}{3})}^2 \frac{( {L_1}^2 + {L_2}^2)}{{L_1}^2 {L_2}^2} \frac{2-s}{2( 6-s)} \\
&& \sim {2-s\over 18(6-s)} {({\frac{R_L}{min[L_1,L_2]}})}^2
\end{eqnarray*} 
\be
\label{equ}
\ee

Perpendicular spatial diffusion is thus 
dominated by turbulent forces at low
particle momenta, where the gyroradius is less than the minimum of the
perpendicular magnetic field focusing lengths. Alternatively, at high
momenta, where the gyroradius is larger than the minimum of the
perpendicular magnetic fierld focusing lengths, perpendicular diffusion
is dominated by the mirror force contribution.

 \section{Summary and conclusions}

In a large-scale magnetized plasma the description of cosmic ray transport is given by the solution of the Vlasov equation for the
particle distribution function. The influence of stocastically fluctuating fields on the particle distribution function can be studied by 
looking for an ensemble-averaged solution of the Vlasov equation, which results from averaging over different realizations of turbulent fields 
with the same statistical properties. 

In the small Larmor approximation it was shown in Paper 1 that one can obtain the solution of the Vlasov equation for 
arbitrary gyrophase motions of the particles,  extending the quasilinear approximation to the particle orbit. 
In Paper 1 the transport parameters of energetic charged particles in turbulent magnetized cosmic 
plasmas were derived for the case of an incompressible plasma, i.e. 
plasmas 
for which the component of the magnetic turbulence, $\delta B_z=0$, parallel to the guide magnetic field, $\vec{B_0}=B_0\vec{e}_{z}$, is set to zero. Here we present 
the generalization of the theory to the case of compressible magnetic turbulence with $\delta B_z\ne 0$. 
Under the assumption that the turbulence is quasi-stationary and homogeneous we have obtained the 
gyro-averaged \fp coefficients for a Corrsin type of generalized orbits. For an axisymmetric turbulence we have derived upper and lower limits for the perpendicular and pitch-angle 
\fp coefficients in the diffusion limit. We have shown upper and lower limits for the perpendicular and parallel spatial diffusion 
coefficients, respectively, describing the spatial diffusion of 
the isotropic part of the \kr phase space density. Finally using the upper and lower limits for the turbulent motion we compare the effects on the transport of cosmic ray particles of the turbulent and 
of mirror forces, if the latter cannot be neglected.

\acknowledgments

This work was partially supported by the German Ministry for Education
and Research (BMBF) through Verbundforschung Astroteilchenphysik grant
05A11PC1 and the Deutsche Forschungsgemeinschaft through grant
Schl 201/23-1.

\section{Appendix: Quasilinear limit}

Following the approach in Paper 1 and assuming $\delta \phi =0$ for the particle orbit in Eq. \ref{b91}
\be
G(\xi ) = \Omega \xi + \delta \phi (\xi )
\ee
we obtain the quasilinear approximation to the particle orbits (Shalchi and Schlickeiser 2004). The argument of 
the Bessel functions of first kind

\be
Z=\kper v\sqrt{1-\mu ^2}
\l[\l(\int^{\tau }d\xi \; \cos (G(\xi ))\r)^2+\l({1\over \Omega }+\int^{\tau }d\xi \; \sin (G(\xi ))\r)^2\r]^{1/2}
\label{b22a}
\ee
at order $n=0$ becomes
\bdm
Z=Z_0={\kper v\sqrt{1-\mu ^2}\over \Omega }\l[\sin ^2(\Omega \tau )+\l(1-\cos ^2(\Omega \tau )\r)^2\r]^{1/2}=
\edm
\be
{\kper v\sqrt{1-\mu ^2}\over \Omega }\l[2\l(1-\cos (\Omega \tau )\r)\r]^{1/2}=
{2\kper v\sqrt{1-\mu ^2}\over \Omega }|\sin ({\Omega \tau \over 2})|
\label{d1}
\ee

The perpendicular \fp coefficients (\ref{c2}) then become 

\bdm
D^{QL}_{ij}=\Re \, \pi v^2\int _0^\infty d\tau \int_{-\infty}^{\infty}d\kpa \int _0^\infty d\kper \, \kper 
e^{\imath \kpa \vpa \tau }J_0\l({2\kper \vper \over \Omega }|\sin \l({\Omega \tau \over 2}\r)|\r),
\edm
\be
\Bigl[ 2 {\mu}^2 P_{ij}(\kpa , \kper , \tau ) +  (-1)^{j \mid i-j \mid}(1-\mu ^2) \cos \l( \mid i-j \mid \frac{\pi}{2} - \Omega \tau \r)  P_{zz} (\kpa , \kper , \tau )  \Bigr],
\label{d21}
\ee

The mixed \fp coefficients (\ref{c22} and \ref{c33}) are 

\bdm
D^{QL}_{i \mu }=\Re \, \pi v \Omega (1-\mu ^2) \int_{-\infty }^\infty d\kpa \, \int_0^\infty d\kper \, \kper 
\int_{0}^{t-t_0}d\tau \, e^{\imath v\mu \kpa \tau } \, J_0\l({2\kper \vper \over \Omega }|\sin \l({\Omega \tau \over 2}\r)|\r)
\edm
\be
\times
\Bigl[ \sin \l( (i-1) \frac{\pi}{2} - \Omega \tau \r) P_{z1}( \kpa , \kper, \tau ) + {(-1)}^{i} \cos \l( (i-1) \frac{\pi}{2} - \Omega \tau \r) P_{z2}(\kpa , \kper , \tau ) \Bigr] 
\label{d22}
\ee

\bdm
D^{QL}_{\mu i }=\Re \, \pi v \Omega (1-\mu ^2) \int_{-\infty }^\infty d\kpa \, \int_0^\infty d\kper \, \kper 
\int_{0}^{t-t_0}d\tau \, e^{\imath v\mu \kpa \tau } J_0\l({2\kper \vper \over \Omega }|\sin \l({\Omega \tau \over 2}\r)|\r)
\edm
\be
\times
\Bigl[ {(-1)}^{i} \sin \l( (i-1) \frac{\pi}{2} - \Omega \tau  \r) P_{1z}( \kpa , \kper , \tau ) -  \cos \l( (i-1) \frac{\pi}{2} - \Omega \tau \r) P_{2z}( \kpa , \kper , \tau 
) \Bigr]
\label{d33}
\ee

whereas the \fp coefficient (\ref{c6}) reduces to  

\bdm
D^{QL}_{\mu \mu }=\Re \, \pi \Omega ^2(1-\mu ^2)\int _0^{t-t_0}d\tau 
\int_{-\infty}^{\infty}d\kpa \int _0^\infty d\kper \, \kper J_0\l({2\kper \vper \over \Omega }|\sin \l({\Omega \tau \over 2}\r)|\r)
\edm
\be
\times
\Bigl[e^{\imath (v\mu \kpa +\Omega )\tau }P_{LL}(\kpa , \kper ,\tau )+e^{\imath (v\mu \kpa -\Omega )\tau }P_{RR}(\kpa , \kper ,\tau )\Bigr]
\label{d3}
\ee

Note that the quasilinear \fp coefficients in axisymmetric turbulence no longer 
involve infinite sums 
of products of Bessel functions which enormously facilitates their numerical computation for specified turbulence field correlation tensors. 

We now explicitly calculate the different \fp coefficients 

\bdm
D^{QL}_{xx}=\Re \, \pi v^2\int _0^\infty d\tau \int_{-\infty}^{\infty}d\kpa \int _0^\infty d\kper \, \kper 
e^{\imath \kpa \vpa \tau }J_0\l({2\kper \vper \over \Omega }|\sin \l({\Omega \tau \over 2}\r)|\r),
\edm
\be
\Bigl[ 2 {\mu}^2 P_{xx}(\kpa , \kper , \tau ) + (1-\mu ^2) \cos \l(  \Omega \tau \r)  P_{zz} (\kpa , \kper , \tau )  \Bigr],
\label{d21m}
\ee

\bdm
D^{QL}_{xy}=\Re \, \pi v^2\int _0^\infty d\tau \int_{-\infty}^{\infty}d\kpa \int _0^\infty d\kper \, \kper 
e^{\imath \kpa \vpa \tau }J_0\l({2\kper \vper \over \Omega }|\sin \l({\Omega \tau \over 2}\r)|\r),
\edm
\be
\Bigl[ 2 {\mu}^2 P_{xy}(\kpa , \kper , \tau ) + (1-\mu ^2) \sin \l( \Omega \tau \r)  P_{zz} (\kpa , \kper , \tau )  \Bigr],
\label{d22m}
\ee

\bdm
D^{QL}_{yx}=\Re \, \pi v^2\int _0^\infty d\tau \int_{-\infty}^{\infty}d\kpa \int _0^\infty d\kper \, \kper 
e^{\imath \kpa \vpa \tau }J_0\l({2\kper \vper \over \Omega }|\sin \l({\Omega \tau \over 2}\r)|\r),
\edm
\be
\Bigl[ 2 {\mu}^2 P_{yx}(\kpa , \kper , \tau ) - (1-\mu ^2) \sin \l( \Omega \tau \r)  P_{zz} (\kpa , \kper , \tau )  \Bigr],
\label{d23}
\ee

\bdm
D^{QL}_{yy}=\Re \, \pi v^2\int _0^\infty d\tau \int_{-\infty}^{\infty}d\kpa \int _0^\infty d\kper \, \kper 
e^{\imath \kpa \vpa \tau }J_0\l({2\kper \vper \over \Omega }|\sin \l({\Omega \tau \over 2}\r)|\r),
\edm
\be
\Bigl[ 2 {\mu}^2 P_{yy}(\kpa , \kper , \tau ) + (1-\mu ^2) \cos \l( \Omega \tau \r)  P_{zz} (\kpa , \kper , \tau )  \Bigr],
\label{d24}
\ee

\bdm
D^{QL}_{x \mu }=\Re \, \pi v \Omega (1-\mu ^2) \int_{-\infty }^\infty d\kpa \, \int_0^\infty d\kper \, \kper 
\int_{0}^{t-t_0}d\tau \, e^{\imath v\mu \kpa \tau } \, J_0\l({2\kper \vper \over \Omega }|\sin \l({\Omega \tau \over 2}\r)|\r)
\edm
\be
\times
\Bigl[ - \sin \l( \Omega \tau \r) P_{zx}( \kpa , \kper, \tau ) - \cos \l( \Omega \tau \r) P_{zy}(\kpa , \kper , \tau ) \Bigr] 
\label{e221}
\ee

\bdm
D^{QL}_{y\mu }=\Re \, \pi v \Omega (1-\mu ^2) \int_{-\infty }^\infty d\kpa \, \int_0^\infty d\kper \, \kper 
\int_{0}^{t-t_0}d\tau \, e^{\imath v\mu \kpa \tau } \, J_0\l({2\kper \vper \over \Omega }|\sin \l({\Omega \tau \over 2}\r)|\r)
\edm
\be
\times
\Bigl[ \cos \l( \Omega \tau \r) P_{zx}( \kpa , \kper, \tau ) + \sin \l( \Omega \tau \r) P_{zy}(\kpa , \kper , \tau ) \Bigr] 
\label{e222}
\ee

\bdm
D^{QL}_{\mu x }=\Re \, \pi v \Omega (1-\mu ^2) \int_{-\infty }^\infty d\kpa \, \int_0^\infty d\kper \, \kper 
\int_{0}^{t-t_0}d\tau \, e^{\imath v\mu \kpa \tau } J_0\l({2\kper \vper \over \Omega }|\sin \l({\Omega \tau \over 2}\r)|\r)
\edm
\be
\times
\Bigl[  \sin \l( \Omega \tau  \r) P_{xz}( \kpa , \kper , \tau ) -  \cos \l( \Omega \tau \r) P_{yz}( \kpa , \kper , \tau ) \Bigr]
\label{e331}
\ee

\bdm
D^{QL}_{\mu y }=\Re \, \pi v \Omega (1-\mu ^2) \int_{-\infty }^\infty d\kpa \, \int_0^\infty d\kper \, \kper 
\int_{0}^{t-t_0}d\tau \, e^{\imath v\mu \kpa \tau } J_0\l({2\kper \vper \over \Omega }|\sin \l({\Omega \tau \over 2}\r)|\r)
\edm
\be
\times
\Bigl[ \cos \l( \Omega \tau  \r) P_{xz}( \kpa , \kper , \tau ) -  \sin \l( \Omega \tau \r) P_{yz}( \kpa , \kper , \tau ) \Bigr]
\label{e332}
\ee

Using the Bessel function addition theorem (see Appendix in Paper 1) 
with $r_1=r_2=1$, $\lambda =\kper \vper /\Omega $ and $\theta =\Omega \tau $ then 

\be
J_0(Z_0)=J_0\l({\kper \vper \over \Omega }\l[2(1-\cos \Omega \tau )\r]^{1/2}\r)
=\sum _{n=-\infty}^\infty J_n^2\l({\kper \vper \over \Omega }\r)e^{\imath n\Omega \tau }
\label{d4}
\ee

so that the perpendicular \fp coefficients can be written as

\bdm
D^{QL}_{xx}=\Re \,\pi v^2 \int _0^{t-t_0} d\tau  \int_{-\infty}^{\infty}d\kpa \int _0^\infty d\kper \, \kper  
\edm
\bdm
[ 2 {\mu}^2 P_{xx}(\kpa , \kper , \tau ) \sum _{n=-\infty}^\infty e^{\imath (\kpa \vpa -n\Omega )\tau  }J_n^2\l({\kper \vper \over \Omega }\r) + \frac{1}{2} (1-\mu ^2) 
\edm
\be
 \sum _{n=-\infty}^\infty e^{\imath (\kpa \vpa -n\Omega )\tau  } \l (J_{n-1}^2\l({\kper \vper \over \Omega }\r) +
J_{n+1}^2\l({\kper \vper \over \Omega }\r) \r)P_{zz}(\kpa , \kper ,\tau )   ],
\label{d51}
\ee

\bdm
D^{QL}_{xy}=\Re \,\pi v^2 \int _0^{t-t_0} d\tau  \int_{-\infty}^{\infty}d\kpa \int _0^\infty d\kper \, \kper  
\edm
\bdm
[ 2 {\mu}^2 P_{xy}(\kpa , \kper , \tau ) \sum _{n=-\infty}^\infty e^{\imath (\kpa \vpa -n\Omega )\tau  }J_n^2\l({\kper \vper \over \Omega }\r) + \frac{1}{2i} (1-\mu ^2) 
\edm
\be
 \sum _{n=-\infty}^\infty e^{\imath (\kpa \vpa -n\Omega )\tau  } \l (J_{n-1}^2\l({\kper \vper \over \Omega }\r) -
J_{n+1}^2\l({\kper \vper \over \Omega }\r) \r)P_{zz}(\kpa , \kper ,\tau )],
\label{d52}
\ee

\bdm
D^{QL}_{yx}=\Re \,\pi v^2 \int _0^{t-t_0} d\tau  \int_{-\infty}^{\infty}d\kpa \int _0^\infty d\kper \, \kper  
\edm
\bdm
[ 2 {\mu}^2 P_{yx}(\kpa , \kper , \tau ) \sum _{n=-\infty}^\infty e^{\imath (\kpa \vpa -n\Omega )\tau  }J_n^2\l({\kper \vper \over \Omega }\r) - \frac{1}{2i} (1-\mu ^2) 
\edm
\be
\sum _{n=-\infty}^\infty e^{\imath (\kpa \vpa -n\Omega )\tau  } \l (J_{n-1}^2\l({\kper \vper \over \Omega }\r) -
J_{n+1}^2\l({\kper \vper \over \Omega }\r) \r)P_{zz}(\kpa , \kper ,\tau ) ],
\label{d53}
\ee

\bdm
D^{QL}_{yy}=\Re \,\pi v^2 \int _0^{t-t_0} d\tau  \int_{-\infty}^{\infty}d\kpa \int _0^\infty d\kper \, \kper  
\edm
\bdm
[ 2 {\mu}^2 P_{yy}(\kpa , \kper , \tau ) \sum _{n=-\infty}^\infty e^{\imath (\kpa \vpa -n\Omega )\tau  }J_n^2\l({\kper \vper \over \Omega }\r)+ \frac{1}{2} (1-\mu ^2) 
\edm
\be
\sum _{n=-\infty}^\infty e^{\imath (\kpa \vpa -n\Omega )\tau  } \l (J_{n-1}^2\l({\kper \vper \over \Omega }\r) -
J_{n+1}^2\l({\kper \vper \over \Omega }\r) \r)P_{zz}(\kpa , \kper ,\tau ) ],
\label{d54}
\ee

Using the addition theorem (\ref{d4}) the pitch angle \fp coefficient (\ref{d3}) becomes 

\bdm
D^{QL}_{\mu \mu }=\Re \, \pi \Omega ^2(1-\mu ^2)\sum_{n=-\infty }^\infty 
\int_{-\infty}^{\infty}d\kpa \int _0^\infty d\kper \, \kper 
\int _0^\infty d\tau  e^{-\imath (n\Omega +\kpa \vpa )\tau }
\edm
\be
\times
\l[J_{n-1}^2\l({\kper \vper \over \Omega }\r)P_{LL}(\kpa , \kper ,\tau )+
J_{n+1}^2\l({\kper \vper \over \Omega }\r)P_{RR}(\kpa , \kper ,\tau )\r],
\label{d6}
\ee
The pitch angle coefficient in (\ref{d6}) does not change with respect to the case of incompressible plasma treated in Paper 1. As remarked in Paper 1 
the pitch angle coefficient agrees exactly with Eq. (12.2.5) of Schlickeiser (2002) for incompressible, axisymmetric turbulence.

\end{document}